\documentclass[10pt, conference, a4paper]{IEEEtran}

\usepackage{url}
\usepackage{graphicx}
\usepackage{amsfonts}
\usepackage{caption}
\usepackage{subcaption}
\usepackage{color}
\usepackage{stfloats}
\usepackage{color,soul}
\usepackage{amsmath}
\usepackage{multirow}
\usepackage{hhline}
\usepackage{textcomp}
\usepackage{gensymb}
\usepackage[ruled,vlined]{algorithm2e}
\usepackage{float}
\usepackage[dvipsnames]{xcolor}
\usepackage{comment}
\usepackage{mwe}
\usepackage[T1]{fontenc}
\usepackage{xcolor}
\usepackage[autostyle]{csquotes}

\usepackage{xcolor,colortbl}
\usepackage{tikz}
\usepackage{array}
\newcolumntype{P}[1]{>{\centering\arraybackslash}p{#1}}

\IEEEoverridecommandlockouts                              

\begin{document}

\title{5G-NIDD: A Comprehensive Network Intrusion Detection Dataset Generated over 5G Wireless Network}

\author{
\IEEEauthorblockN{
Sehan Samarakoon\IEEEauthorrefmark{1},
Yushan Siriwardhana\IEEEauthorrefmark{2},
Pawani Porambage\IEEEauthorrefmark{3},
Madhusanka Liyanage\IEEEauthorrefmark{4},
Sang-Yoon Chang\IEEEauthorrefmark{5},\\
Jinoh Kim\IEEEauthorrefmark{6},
Jonghyun Kim\IEEEauthorrefmark{7},
Mika Ylianttila\IEEEauthorrefmark{8}}

\IEEEauthorblockA{\IEEEauthorrefmark{1}\IEEEauthorrefmark{2}\IEEEauthorrefmark{3}\IEEEauthorrefmark{4}\IEEEauthorrefmark{8}Centre for Wireless Communications, University of Oulu, Finland}
\IEEEauthorblockA{\IEEEauthorrefmark{3}VTT Technical Research Centre, Finland}
\IEEEauthorblockA{\IEEEauthorrefmark{4}School of Computer Science, University College Dublin, Ireland}
\IEEEauthorblockA{\IEEEauthorrefmark{5}Department of Computer Science, University of Colorado Colorado Springs, USA}
\IEEEauthorblockA{\IEEEauthorrefmark{6}Computer Science Department, Texas A\&M University--Commerce, USA}
\IEEEauthorblockA{\IEEEauthorrefmark{7}Information Security Research Division, ETRI (Electronics and Telecommunications Research Institute), Korea}
\IEEEauthorblockA{Email: \IEEEauthorrefmark{1}\IEEEauthorrefmark{2}\IEEEauthorrefmark{4}\IEEEauthorrefmark{8}[firstname.lastname]@oulu.fi, 
\IEEEauthorrefmark{3}pawani.porambage@vtt.fi,
\IEEEauthorrefmark{4}madhusanka@ucd.ie,
\IEEEauthorrefmark{5}schang2@uccs.edu,\\
\IEEEauthorrefmark{6}jinoh.kim@tamuc.edu
\IEEEauthorrefmark{7}jhk@etri.re.kr}
}

\maketitle

\begin{abstract}
With a plethora of new connections, features, and services introduced, the 5th generation (5G) wireless technology reflects the development of mobile communication networks and is here to stay for the next decade. The multitude of services and technologies that 5G incorporates have made modern communication networks very complex and sophisticated in nature. This complexity along with the incorporation of Machine Learning~(ML) and Artificial Intelligence (AI) provides the opportunity for the attackers to launch intelligent attacks against the network and network devices. These attacks often traverse undetected due to the lack of intelligent security mechanisms to counter these threats. Therefore, the implementation of real-time, proactive, and self-adaptive security mechanisms throughout the network would be an integral part of 5G as well as future communication systems. Therefore, large amounts of data collected from real networks will play an important role in the training of AI/ML models to identify and detect malicious content in network traffic. This work presents 5G-NIDD, a fully labeled dataset built on a functional 5G test network that can be used by those who develop and test AI/ML solutions. The work further analyses the collected data using common ML models and shows the achieved accuracy levels.

\end{abstract}

\begin{IEEEkeywords}
5G Security, 5GTN, Dataset, Intrusion Detection, Machine Learning
\end{IEEEkeywords}

\section{Introduction}
\label{sec:intro}

The advancements with 5th generation (5G) wireless systems increase the complexity of wireless networks in terms of heterogeneity, device count, and the amount of data generated. Beyond 5G, the 6G era drives a revolution of wireless networks from connected things to connected intelligence~\cite{letaief2019roadmap}, leading towards more complex networks. Artificial Intelligence~(AI) and Machine Learning ~(ML) play a substantial role in beyond 5G networks in many aspects including intelligent wireless communications, closed-loop optimization of networks, and big data analytics. The 6th generation (6G) era envisions the adoption of network-wide AI~\cite{siriwardhana2021ai}. As the complexity of the networks grows higher, the threat surface also becomes broader due to the heterogeneity, massive scale, and the plethora of applications served by these networks~\cite{saad2019vision}.

Security of 5G and 6G networks is of the utmost importance as the novel applications demand extreme requirements from the networking infrastructure~\cite{porambage2021roadmap}. The potential of adversaries uncovering network vulnerabilities using AI techniques increases the challenge of designing proper security mechanisms to protect the network. Conventional reactive security approaches where the resolution action starts after the attack is detected, are insufficient against such intelligent attacks~\cite{9815812}. Therefore, a paradigm of proactive, self-aware, and self-adaptive intelligent security systems is a must~\cite{siriwardhana2021ai}. Proper use of massive data generated by future networks leads to uncovering anomalous network behaviors, hence providing valuable inputs for the designing of intelligent security mechanisms. Therefore, it is evident that the design and implementation of AI-based intrusion detection and prevention approaches are much needed to secure future networks~\cite{santos2020machine}, complementing and improving the present security designs.

The majority of data generated inside the network is extremely important as AI-based security mechanisms are capable of uncovering hidden patterns in large volumes of data. A prevailing problem in AI-based security research is the unavailability of comprehensive valid datasets that exerts complex network behaviors~\cite{moustafa2015unsw}, mainly 5G network behaviors. In the regime of supervised learning, the training datasets are scarce, which prevents the accurate evaluation of novel AI-based security techniques for future networks. Some of the well-known publicly available datasets are outdated~\cite{bhuyan2015towards} and have limited applicability in 5G network security research. Mobile Network Operators~(MNO) are usually reluctant to publicly share the data from their live network traffic due to privacy issues, and the potential threats of exposing network vulnerabilities to external parties. Hence, a comprehensive dataset from an actual running network that exerts 5G behavior has substantial benefits in AI-based 5G security research and those who test AI/ML algorithms.

This article publishes 5G-NIDD, a network intrusion detection dataset generated from a real 5G test network. To the best of our knowledge, there is no prior work in the state-of-the-art that uses 5G network flows to create a dataset. Therefore, 5G-NIDD contains unique features that exist in 5G network flows compared with the past datasets. Usually, many available datasets are sourced from either virtualized networks or temporary networks specifically designed to create a dataset. Even though the simulators and virtual environments are frequently utilized for research and development, real-world settings and testbeds are necessary to assess the performance of new applications and services before deployment~\cite{piri20165gtn}. Aligning with that concept, 5G-NIDD is generated using a fully functional 5G network created for 5G testing purposes, thus providing a close resemblance to a real network scenario. The dataset is collected using the 5G Test Network (5GTN) in Oulu, Finland \cite{5gtnf}. 5GTN serves as the core network and offers application developers a carrier-grade test network by deploying a variety of measuring and monitoring tools throughout the network~\cite{piri20165gtn}. 5G-NIDD presents a combination of attack traffic and benign traffic under different attack scenarios. We use real mobile devices attached to the 5GTN to generate benign traffic whereas in some other datasets pre-collected data has been added as benign traffic after post-processing. In particular, our contributions to this paper are as follows. 

\begin{itemize}
\item Implement the whole network setup using appropriate hardware and software on a real 5G network environment and configure the roles and positions of attackers/targets to simulate different attack scenarios.
  \item Create and publish 5G-NIDD, a labeled network intrusion detection dataset that contains nine intrusion types along with benign network traffic. The dataset allows researchers to test novel intrusion detection algorithms. 
  \item Provide extensive evaluation of 5G-NIDD using multiple ML techniques, present the accuracy levels and the validity of the dataset in network intrusion detection.
\end{itemize}

The remainder of the paper is organized as follows: Section \ref{sec:relatedwork} explains the background and related work including past datasets and their shortcomings. Section \ref{sec:generation} presents details of the testbed attached to 5GTN which is used for the dataset generation. Section~\ref{sec:attacks} describes the attack scenarios, attack tools, and the methodology we followed to create the dataset. Section \ref{sec:dataset} explains the details of the post-processing of data. Section \ref{sec:analysis} discusses the results obtained using different ML models to test the dataset. Section \ref{sec:futurework} provides potential expansions of the testbed and future directions to create more sophisticated datasets. Finally, Section \ref{sec:conclusions} concludes the paper.

\section{Background and Related Work}
\label{sec:relatedwork}

This section presents a summary of the commonly available threat detection mechanisms of networks and explains why those threat detection mechanisms are inefficient in 5G and 6G networks. It also elaborates the behavioral-based anomaly detection equipped with AI to complement the existing mechanisms. This section also summarizes some of the available datasets for ML-based intrusion detection.

\subsection{Conventional Threat Detection}

Conventional threat detection involves the use of knowledge-based or signature-based intrusion detection techniques. These signature-based threat detection systems use the available signatures which are stored in a database for pattern matching to detect the known threats \cite{khraisat2018anomaly}. The database keeps updating its signatures based on the new attacks. If a signature of a specific attack is missing in the database, the system fails to detect that as an attack. This detection scheme is not sufficient for 5G and beyond networks as the likelihood of occurring intelligent zero-day type attacks is high~\cite{khraisat2019survey}. Furthermore, signature-based Intrusion Detection Systems (IDS) perform poorly when the attack spans across multiple numbers of packets as only individual packets are compared against the signatures database in general. The comparison of each packet with the signature database is expensive in terms of time and resources. This is intolerable in a real-time 5G network with applications demanding ultralow latencies. 

\subsection{ML based Threat Detection}

ML based threat detection approaches have substantial popularity with the advancement of AI techniques~\cite{9815754}. Salo et al. performed a literature review to recognize 19 different methods for data mining that are frequently used in network intrusion detection~\cite{salo2018data,salo2020data}. Subsequently, they proposed an ensemble feature selection and anomaly detection mechanism, highlighting the need for more ML-based research to implement real-time intrusion detection~\cite{salo2019clustering}. Li et al. proposed a decision tree based intrusion detection system for the connected and autonomous vehicles to identify both intra and inter-vehicle network attacks~\cite{yang2019tree}. Verma et al. performed k-nearest neighbor classification and k-means clustering techniques to measure the performance of CIDDS-001 dataset using different evaluation metrics~\cite{verma2018statistical}. The work concludes that the two techniques are effective to use in anomaly detection in flow based CIDDS-001 dataset.
Gumus et al. proposed an online implementation of Naive Bayes classifier with moving average and standard deviation. Injadat et al. proposed a multistage optimized ML based Network Intrusion Detection System (NIDS) framework to lower computing complexity while retaining detection performance~\cite{injadat2020multi}. The work compared the effect of oversampling techniques on training sample size and the effect of different feature selection techniques such as information gain and correlation based methods~\cite{injadat2020multi}.

Self-adaptive ML based anomaly detection in 5G networks~\cite{maimo2018self} utilizes AI algorithms to identify cyber threats and automatically adapts to optimize computing resources based on the traffic flow. These detection schemes use a large amount of data to learn and then apply the learned knowledge to identify traffic flows that deviate from normal traffic behaviors. The possession of high computational ability at commendable speeds in today's 5G networks allows the proper use of AI/ML techniques in intrusion detection. AI/ML techniques have the potential to deliver real-time solutions in detecting and neutralizing more sophisticated attacks in 5G and beyond networks.

AI/ML-based threat detection schemes possess the ability to learn and enhance their detection capability from experience without a specific need for programming~\cite{saranya2020performance}. Several categories of AI-based techniques exist based on the way learning occurs, including supervised learning, unsupervised learning, semi-supervised learning, and reinforcement learning~\cite{sarker2021machine}. Supervised learning uses labeled data to reveal a relationship between inputs and outputs and predicts the output for a completely new set of inputs. Well-known supervised learning algorithms/models include Decision Tree, Random Forest, Naive-bayes, and deep neural networks. The presence of a labeled dataset with a sufficient amount of data is important in supervised learning to train a good model.

\subsection{Existing Datasets}

The first widely known dataset for ML based intrusion detection is DARPA dataset~\cite{lippmann2000evaluating, darpa98} which consists of simulated Denial of Service (DoS) attacks, guess password, buffer overflow, synflood, and namp attacks. Research communities developed more datasets to support ML based intrusion detection to address the weaknesses of existing datasets by providing further improvements. Examples for such datasets include KDD Cup 99~\cite{kdd99}, NSL KDD~\cite{nslkdd}, DEFCON~\cite{defcon}, CAIDA~\cite{caida}, LBNL~\cite{nechaev2004lawrence}, CTU-13~\cite{ctu-13}, UNSW-NB 15~\cite{moustafa2015unsw}, and Bot-IoT~\cite{koroniotis2019towards} datasets. A recent survey briefly explains a comprehensive summary of such datasets until 2020 that are effective to evaluate intrusion detection on networks~\cite{shaukat2020survey}. The majority of these datasets are outdated for modern networks as they are compiled before some of the drastic technological evolutions. However, the CICIDS2017 and Bot-IoT datasets are more recent datasets. The CICIDS2017 dataset~\cite{sharafaldin2018toward} contains multiple attack types including DoS, Distributed Denial of Service (DDoS), Heartbleed, and infiltration attacks. The dataset is based on a realistic network that combines an attack network and a victim network. The schedule for different attacks spans a period of four days. The benign traffic profile contains Hypertext Transfer Protocol (HTTP), Hypertext Transfer Protocol Secure (HTTPS), File Transfer Protocol (FTP), Secure Shell Protocol (SSH), and email protocols to simulate the behavior of a real network. The Bot-IoT dataset~\cite{koroniotis2019towards} proposed in 2018 consists of more than 72 million records. This dataset consists of DoS attacks, probing attacks, and information theft in an IoT network simulated using Node-red tool. The IoT devices include a weather station, a smart fridge, motion-activated lights, a remotely activated garage door, and a smart thermostat. The entire dataset was created using a virtual environment which is created inside one physical computer where the attackers and the victims are created as virtual instances. Some of the datasets were subjected to different critics such as the presence of artificial traffic, higher amounts of redundancy, ignorance of real-world conditions, lack of diversity, and the presence of highly anonymized content in packets~\cite{kdd99}-~\cite{sharafaldin2018toward}.

\begin{table*}[htbp]
\caption{Existing Intrusion Detection Datasets}
    \label{tab:datasets}
    \centering
    \begin{tabular}{|m{1.7cm}|m{3.5cm}|m{3.6cm}|m{3cm}|m{3.6cm}|} 
    \hline
    Dataset  & Scenario/application & Data production method & Attack types & Limitations/issues \\ \hline \hline
       DARPA~\cite{darpa98} & Computer network intrusion detection system~\cite{yavanoglu2017review} & Transmission Control Protocol (TCP) dump and log data~\cite{thomas2008usefulness} & DoS, probing, privilege escalation & Usage of artificially simulated data for background traffic~\cite{thomas2008usefulness}\\ \hline
       
       KDD Cup 99~\cite{kdd99} & Computer network intrusion detection system & Generated by processing the tcpdump portion of DARPA dataset~\cite{tavallaee2009detailed} & DoS, probing, privilege escalation & Large number of redundant records, data corruptions which led to skewed testing results~\cite{tavallaee2009detailed} \\ \hline
       
       NSL KDD~\cite{nslkdd} & Computer network intrusion detection system & Selected records of KDD 99 dataset by removing redundant records~\cite{tavallaee2009detailed} & DoS, probing, privilege escalation & Does not represent modern low foot print attack scenarios~\cite{tavallaee2009detailed}\\ \hline
       
       DEFCON~\cite{defcon} & Traffic produced during a capture the flag competition & Captured in packet based format~\cite{ring2019survey} & Port Scanning, buffer overflow attacks & Traffic different from real world network traffic \\ \hline
       
       CAIDA DDoS 2017~\cite{caida} & One hour DDoS attack traffic~\cite{bhuyan2015towards} & Traffic traces in pcap format~\cite{bhuyan2015towards} & DDoS & Datasets are very specific to particular events and are anonymized with their payload, protocol information and destination~\cite{shiravi2012toward} \\ \hline
       
       LBNL~\cite{nechaev2004lawrence} & Network traffic within enterprise networks~\cite{pang2005first} & Incoming, outgoing and internally routed traffic streams at the LBNL edge routers in packet based format~\cite{bhuyan2015towards} & Port scans & Consists of anonymized traffic with only header data~\cite{dwibedi2020comparative,nechaev2004lawrence} \\ \hline
       
       CTU-13~\cite{ctu-13} & Real mixed botnet traffic~\cite{yavanoglu2017review} & Captured in Pcaps and then converted to NetFlows, WebLogs etc.~\cite{garcia2014ctu}& Botnets~\cite{garcia2014empirical} & Higher percentage of unknown background traffic and complete traffic captures not available publicly~\cite{nilua2019machine}\\ \hline

       UNSW-NB 15 & Computer network security dataset~\cite{zoghi2021unsw} & Used IXIA PerfectStorm tool to produce modern normal and abnormal traffic in pcap and argus file formats~\cite{moustafa2015unsw} & DoS, backdoors, exploits, fuzzers, port scans, worms & The dataset was designed based on a synthetic environment for generating attack activities. Class imbalance and class overlap~\cite{zoghi2021unsw}\\ \hline
       
       CICIDS 2017/2018~\cite{sharafaldin2018toward} & IDS dataset with common updated attacks & Captured in packet based format and converted to bi-directional flow based format~\cite{sharafaldin2018toward} & DoS, DDoS, heartbleed, infiltration, Structured Query Language(SQL) injection, SSH brute force & Many redundant records and high class unbalance~\cite{panigrahi2018detailed} \\ \hline
       bot-IoT~\cite{koroniotis2019towards} & Emulation of both IoT and a botnet~\cite{peterson2021review} & Captured in PCAPS and converted to flows~\cite{peterson2021review} & DoS, probing, information theft & Odd composition of normal traffic simulated from softwares~\cite{peterson2021review}\\ \hline 

       IoT-NI~\cite{q70p-q449-19} & Network attacks in an IoT setting  & Captured in 42 PCAP files under the monitor mode of a wireless network adaptor & DoS, Man In The Middle (MITM), host/port scan, Mirai & Small-scale data collection using two home devices (speaker and camera) as victims \\ \hline

       IoT-23~\cite{iot23} & Network traffic captures under 23 scenarios & Captured in PCAP files with the associated IDS log (created by Zeek~\cite{Bro}) & DDoS, C\&C, FileDownload, HeartBeat, port scan, Botnet (Mirai, Okiru, Torii) & Small-scale collection with three types of home devices (speaker, bulb, door lock) and a high degree of class imbalance \\ \hline

       MQTT-IoT-IDS2020~\cite{hindy2021machine} & Four attack scenarios on a simulated network with 12 sensors & Captured in PCAP files with the corresponding packet- and flow-level feature set & Scan, password cracking & Simple brute force-mannered attacks included in the attack scenarios \\ \hline
    \end{tabular}
    
\end{table*}

\subsection{Limitations in Existing work}

The accuracy and the efficiency of ML-based intrusion detection heavily depend on the quality of the dataset and how close the behavior of the data is to the behavior of a real network scenario. Characteristics of a comprehensive and valid dataset include attack diversity, anonymity, available protocols, complete capture, complete interaction, complete network configuration, complete traffic, feature set, heterogeneity, labeling, and metadata~\cite{gharib2016evaluation}. Recent datasets such as CICIDS2017 exhibit these characteristics and have proven successful in ML based intrusion detection research. However, the behavior of 5G and beyond networks is far from the testbeds or the simulation platforms used to create these datasets. Moreover, there are few publicly available datasets from real MNO networks. Table \ref{tab:datasets} presents a summary of existing intrusion detection datasets along with their limitations. At the time of this writing, there is no suitable dataset generated using a real 5G network, which affects the AI-based intrusion detection research in 5G networks. The objective of this effort is to bridge that gap and create a novel dataset from a real 5G network implementation and publish it for the use of further research.

\section{Dataset Gneration Environment}
\label{sec:generation}

This section extensively describes the environment we used for the generation of 5G-NIDD. It provides detailed information about the 5G Test Network~(5GTN), which is the underlying infrastructure. Then it describes the additional testbed elements added to 5GTN for the purpose of creating the dataset and collecting the data. Finally, it provides details on benign user behavior and benign traffic extraction.

\subsection{Testbed}

\subsubsection{5GTN}

\begin{figure*}[ht]
    \centering
    \includegraphics[width=0.85\textwidth]{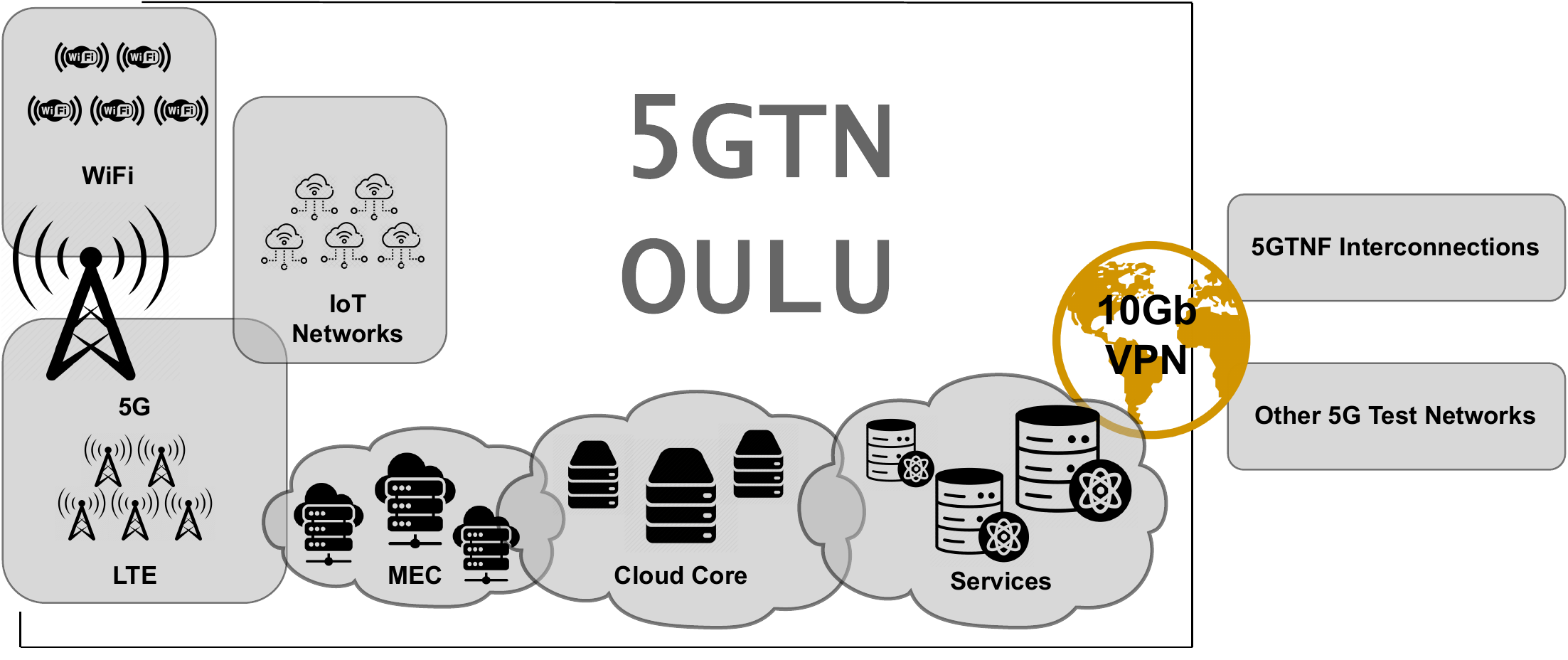}
    \caption{Architecture of 5GTN Oulu \cite{piri20165gtn}.}
    \label{fig:5gtnarch}
\end{figure*}

5G Test Network Finland~\cite{5gtnf} is an open and evolving innovation ecosystem supporting 5G and beyond technology research. It offers a leading edge environment to develop 5G and beyond, AI and cyber security based vertical industry solutions, services, systems and products. 5GTN deploys a number of measurement and monitoring tools in various network areas with the goal of providing application developers with a carrier quality test network~\cite{piri20165gtn}. We used the University of Oulu site of 5GTN for the dataset creation, which is an open test environment as shown in Fig. \ref{fig:5gtnarch}. The deployment of 5GTN used for the dataset creation has the Non-Standalone~(NSA) option 3a~\cite{agiwal2021survey}.

\subsubsection{Test Network}

\begin{figure*}[ht]
    \centering
    \includegraphics[width=0.95\textwidth]{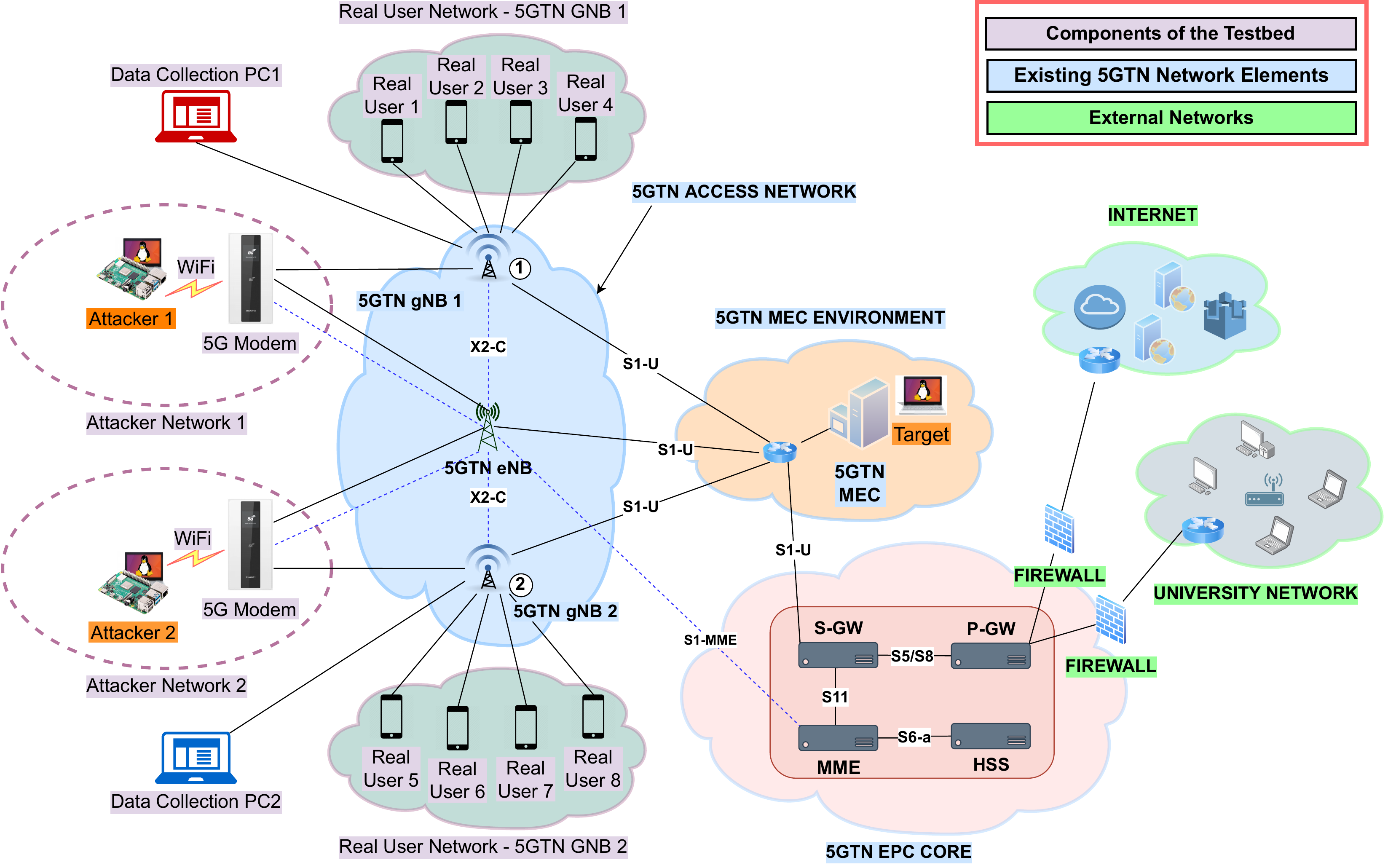}
    \caption{Network architecture of the testbed.}
    \label{fig:networkarch}
\end{figure*}

Fig. \ref{fig:networkarch} depicts additional testbed elements we used along with the 5GTN for the purpose of data collection. We used two Nokia - Flexi Zone Indoor Pico Base Stations existing in 5GTN in our activity. We connected an attacker node and a set of benign traffic-generating devices to of these two base stations. A Dell - N1524 switch connects the attackers to the victim's network. The pico base stations are also connected to the eNodeB through the x2-c interface while the connection between the pico base station and the switch is through S1-U interface.

\subsubsection{Attacker Preparation}

\begin{figure}[ht]
    \centering
    \includegraphics[width=0.48\textwidth]{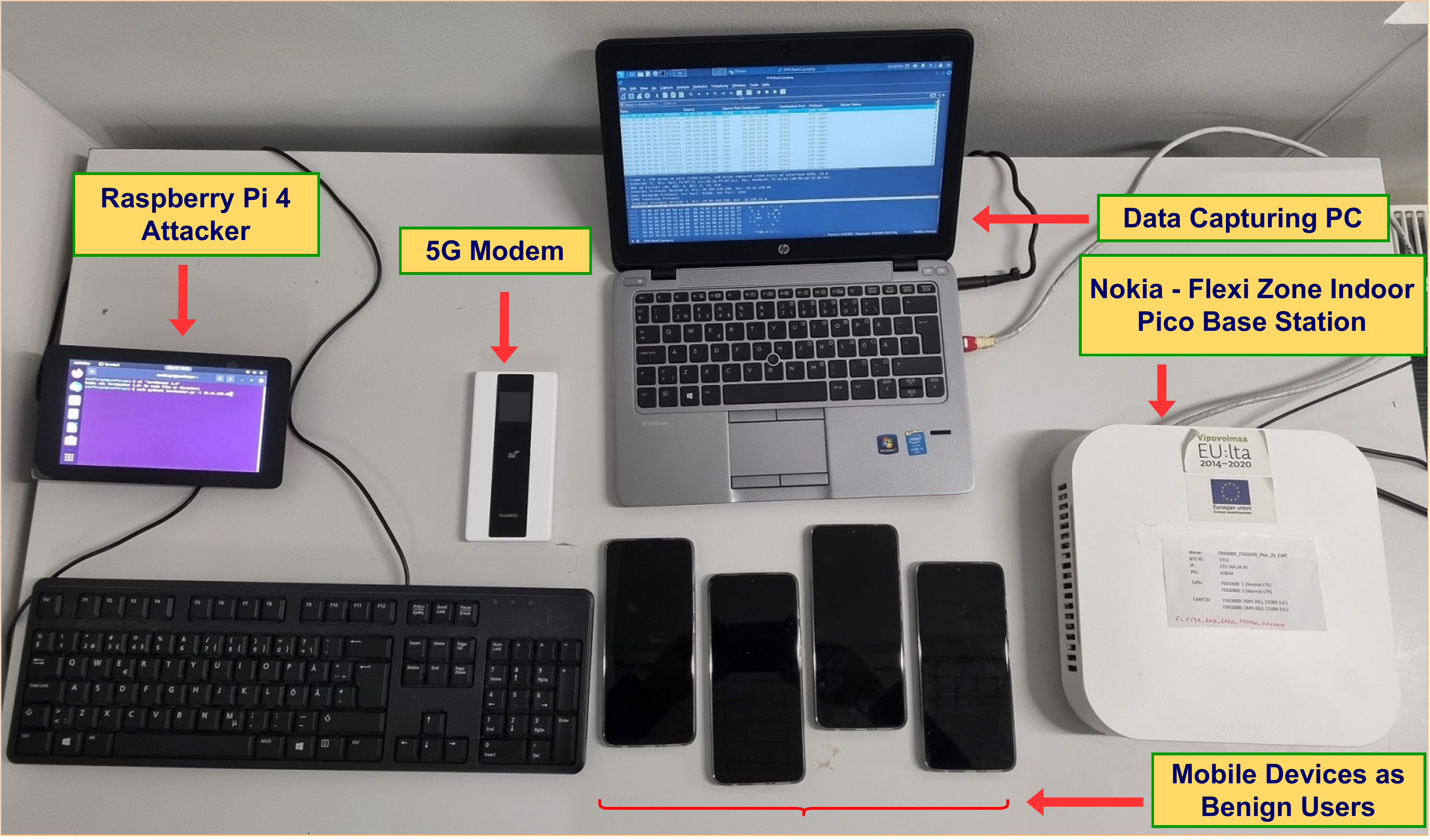}
    \caption{Equipment setup for one base station.}
    \label{fig:testbedpic}
\end{figure}

We configured Raspberry Pi 4 Model B devices as the attacker nodes. We set up Raspberry Pi 4 devices with Ubuntu operating system for easy use of open-source attacking tools. We connected the attackers to the pico base stations via Huawei E6878-870 5G modems. We set up the connection between the attacker node and the modem via WiFi and the connection between the modem and the base station was 5G. Fig. \ref{fig:testbedpic} shows the actual equipment setup for one base station. The setup includes the base station, attacker, benign users, and the data capturing Personal Computer (PC). Table \ref{tab:networkdevices} provides a list of devices used in the testbed.

\subsubsection{Victim Placement}

The victim/target was an Ubuntu server instance located in the Multi-access Edge Computing (MEC). MEC plays an important role in modern 5G networks. MEC facilitates essential requirements in 5G networks in terms of low latency and high bandwidth~\cite{porambage2018survey}. 
We wanted to have the attackers and the target in different networks, therefore we placed the target in the MEC. To launch a successful attack, the attacker needs to send the attack traffic through the 5G network. This makes the deadset have more realistic flows. In 5G networks, attacker presence can be seen in different subnetworks.

\begin{table}[ht]
\caption{Devices of the Testbed}
    \label{tab:networkdevices}
    \centering
    \begin{tabular}{|c|c|} \hline
    Device   &  Description \\ \hline \hline
       Attacker  &  Raspberry Pi 4 Model B with Ubuntu 21.10 - Impish\\ \hline
       Victim & HP with Ubuntu 18.04.5 LTS - Bionic  \\ \hline 
       Base Station & Nokia - Flexi Zone Indoor Pico Base Station  \\ \hline
       Modem & HUAWEI - E6878 \\ \hline
       Mobile devices & Motorola - Moto g50 5G\\ \hline
       Switch & Dell - N1524   \\ \hline
    \end{tabular}
\end{table}

\subsection{Benign Traffic Generation}

A key feature that distinguishes 5G-NIDD from other datasets is the approach for benign traffic generation. Actual mobile devices in the network generate benign traffic instead of simulated traffic. Adding prior simulated benign traffic has been the common approach in most of the other datasets, which affects the traffic behavior in the dataset. We captured the live traffic from mobile devices in the network, which includes both the attack traffic and the benign traffic. The benign traffic profile includes protocols such as HTTP, HTTPS, SSH, and Secure File Transfer Protocol (SFTP). The sources for HTTP and HTTPS are live streaming services and web browsing. We deployed SSH clients and servers on mobile devices, generated SSH and SFTP traces, and added them to the dataset to increase variety.

\section{Attack Types and Attacking Tools}
\label{sec:attacks}

This section describes the attack types and attacking tools that we considered for the generation of 5G-NIDD. As described below, mainly we evaluated for few variants of DoS attacks and Port Scan attacks.  

\subsection{DoS Attacks}

A DoS attack is one of the most common attacks that can occur in a network. The aim of this attack is to slow down or completely shut down a target making it inaccessible to legitimate users. Flooding techniques and malicious content injection are some of the common methods to lunch a DoS attack. The network or a node will eventually undergo resource exhaustion and deny the nonmalicious users' access as a result of a DoS attack. Due to the heterogeneous nature of the 5G networks, DoS attacks impose a vital threat in 5G which may target the network nodes, devices, and applications~\cite{alliance20165g}. This work focuses on the DoS/DDoS attacks targeted at user devices, but the attack may congest the network too. The non-guaranteed availability of security infrastructure for operating systems, applications, and configuration data on user devices requires an adequate level of protection from the networking infrastructure~\cite{ahmad2018overview}. Therefore, intrusion detection mechanisms implemented at the network level is important to detect these type of attacks.

Three most common categories of DoS/DDoS attacks are volume-based, protocol-based, and application layer attacks \cite{singh2017application}. In volume-based DoS attacks, the attacker directs large volumes of attack traffic toward the target to deplete the network resources of the targeted server. Attackers exploit the weaknesses and deficiencies present in the protocols and launch protocol based DoS attacks. Moreover, DoS/DDoS attacks aimed at application layer services are also common. Our work represents all these categories of DoS/DDoS attacks.

\subsubsection{ICMP Flood}

Internet Control Message Protocol (ICMP) flood attacks use ICMP echo requests to flood a target at a very high frequency. Ideally, an ICMP echo reply follows the echo request to the same Internet Protocol (IP) address. In this type of attack, the target will send back echo replies to massive amounts of ICMP echo requests. The higher frequency of echo requests also leads to a higher frequency in the rate of response which eventually overwhelms the network and makes the services unavailable for normal users. We performed the ICMP flood attack with the Hping3 tool using the following command.\\

\textit{sudo hping3 --flood --rand-source -1 -p \textquote{Targeted port/Range of ports} \textquote{Target IP address}}\\

\subsubsection{UDP Flood}

In a User Datagram Protocol (UDP) flood attack, attacker sends UDP packets at a higher rate. Since UDP is a connection-less protocol, it is possible to send a large amount of traffic. As the UDP packets arrive at the target destination, the reply is a \textquotedblleft Destination Unreachable\textquotedblright~packet if no associated application is found. Over time, as a higher amount of UDP packets are received and responded to, the system eventually becomes nonresponsive. We used the same Hping3 tool to perform the UDP flood attack by launching the following command.\\

\textit{sudo hping3 --flood --rand-source --udp -p \textquote{Targeted port/range of ports} \textquote{Target IP address}}\\

\subsubsection{SYN Flood}

SYN flood attack is a result of the exploitation of the TCP three way handshake. In a typical TCP communication between two nodes, the source nodes sends a SYN packet to the desired node for the initiation of the connection. The destination node replies with a SYN ACK to acknowledge the SYN packet. Finally, the source sends another ACK to acknowledge receiving the SYN ACK, completing the three way handshake. In a SYN flood attack, the attacker does not perform the final step of the TCP handshake, therefore leaving the connection half open. The transmission of SYN packets at a higher frequency will leave large amounts of ports half open for a certain period and eventually the receiver will be exhausted once all ports are occupied preventing access for legitimate users. We performed the SYN flood attack using the Hping3 tool through the following command.\\

\textit{sudo hping3 -S -p \textquote{Targeted port/range of ports} --flood --rand-source \textquote{Target IP address}}\\

\subsubsection{HTTP Flood}

HTTP flood attacks target the application layer. It is a popular method for DoS/DDoS attacks due to its proven effectiveness in mimicking normal human behavior and thereby staying undetected. We used Python based Goldeneye tool to conduct the HTTP flood attacks. To perform the application layer attacks, we deployed a web server at the victim MEC instance. The target was an Apache2 web server deployed in MEC. We used the following command to executes the attack.\\

\textit{sudo python3 ./goldeneye.py http://\textquote{Target IP address of the web server}}\\

\subsubsection{Slowrate DoS}

Another exploitation of the application layer to conduct DoS attacks is the use of slow rate attacks. In terms of the speed and the number of packets directed at the target, it is comparatively different from other forms of DoS attacks. This slow nature of the traffic makes it harder to detect. In this exercise, we conducted two different forms of slow rate DoS attacks. As the first one, we performed slowloris attack through a python script. The attack establishes several connections with the targeted web server and maintain them for a longer period. We performed this by sending HTTP partial headers continuously to keep the connections open without completion~\cite{calvert2019impact}. The web server waits for the connections to complete which might never occur. The opening of a large number of such connections will eventually lead to resource exhaustion. We executed the slowloris attack using the following syntax from the terminal.\\

\textit{python3 slowloris.py \textquote{Target IP address of the web server}}\\

In another case, the attacker sends slow POST requests to a web server. The POST requests comprises packets that specify a larger packet size in the header field, but the actual packet has a lesser size. The server then waits for the completion of the whole packet size specified in the header field~\cite{calvert2019impact}. We carried out this specific type of slow DoS attack using the tool Torshammer. We executed the following command from the terminal to perform the attack. \\

\textit{python2 torshammer.py -t \textquote{Target IP address of the web server}}\\

\subsection{Port Scans}

A port scan usually precedes an actual attack to identify opportunities for attacks via open ports. Typically, port scans send requests to a targeted range of ports in a targeted host and observe the response. The response is sufficient to determine the status of a specific port in most cases. In some instances, deeper knowledge such as the operating system of the target may also be discovered~\cite{bhuyan2011surveying}. Port scan is an efficient way to determine exploitable hosts in a network before the execution of attacks. The dataset contains attack data from different port scans such as the SYN scan, TCP Connect scan, and UDP scan. A brief explanation of each of these attacks is given below.

\subsubsection{SYN Scan}

The SYN scan uses part of the three way handshake in TCP connection establishment to discover open ports. The attacker sets the SYN flag in a TCP packet and sends to the target. If the targeted port/s is ready for a TCP connection, it returns a response packet with SYN and ACK flags set. This means the scan indicates an \textit{open} port to the attacker. Ports that are not ready to establish a TCP connection send a RST packet indicating a \textit{closed} port. As the attacker determines the state of the port from two packet flows, attacker cease the execution of the next step of the three way handshake. However, to terminate the connection the attacker may send an RST packet to the target once the information on the status of the port is collected. If not, the target will keep sending packets with SYN and ACK flags set until a response is seen from the attacker. Other than \textit{open} or \textit{closed} ports, a \textit{filtered} port may imply that a firewall is blocking the packets or the host is not reachable.

The incomplete three way handshake in SYN scan makes the scanning process quite fast in comparison to other available techniques. The SYN scan is one of the most popular scanning techniques. We performed the SYN scan through the Nmap open source tool in this work. We specified the targeted port range and the IP address using the following command which executed through the terminal.\\

\textit{sudo nmap -sS \textquote{Target IP address} -p \textquote{Targeted port/range of ports}}\\

\subsubsection{TCP Connect Scan}

Similar to SYN scan, the TCP connect scan exploits the TCP handshake to perform the scanning process. However, the three way handshake is fully completed therefore, the time taken to scan the ports is higher compared with SYN Scan. Conversely, attackers may not need the same level of privileges to execute a TCP connect scan as they require for SYN scan. 

The execution of the TCP connect scan is similar to SYN scan with the addition of the completion step in the three way handshake. This means that the two initial packet flows (SYN flag set packet and SYN ACK flags set packet) take place for an \textit{open} port. Subsequently, the target replies with a packet set with ACK flag to complete the three way handshake. Although the opening of a TCP connection also offers the potential for data exchange, the attacking operating system makes sure the connection is terminated as soon as it detects that one has been made. In the event of a \textit{closed} port, target sends a RST packet to the host similar to the SYN scan. \textit{Filtered} ports may also exist in this scenario. We executed the TCP connect scan through the nmap tool using the following command.\\

\textit{sudo nmap -sT \textquote{Target IP address} -p \textquote{Targeted port/range of ports}}\\

\subsubsection{UDP Scan}

UDP scan transmits UDP datagrams to the targeted ports of the targeted host. The attacker sends the UDP packets with a missing payload to ports that are not UDP. For UDP ports, a protocol-specific payload will exist. In UDP scan, the status of a port is defined based on the response of the target or the unresponsiveness of the target. As nonUDP ports receive a UDP datagram without a payload, the probability of getting a proper response is minimum. An \textit{open|filtered} state is likely from such ports which indicates either a firewall is blocking the packets or packets are simply been forwarded from TCP/IP stack towards a listening application and discards them as invalid. Because of this, it might be challenging to precisely identify whether a port is open for nonUDP ports. In the case of UDP ports, a response may be recognized as a clear indicator of an \textit{open} port. A port is judged to be closed if the response is an ICMP port unreachable error. We used the nmap tool to conduct UDP scan through the following command on the terminal.\\

\textit{sudo nmap -sU \textquote{Target IP address} -p \textquote{Targeted port/range of ports}}\\

\section{Data collection and processing}
\label{sec:dataset}

This section provides a detailed explanation of the schedule of various attacks, data capturing procedures, and the subsequent post-processing of captured data. This includes actions like removing the General Packet Radio Service Tunnelling Protocol (GTP) layer, converting pcap data to a network flow-based format, and feature selection. After the post-processing, the data is ready to feed ML algorithms.

\subsection{File Capture}

We performed the data collection process on two different days and captured both attack and benign traffic passing through the network at the two base stations.

We captured the data for each attack in different sessions in the pcap format. For port scans, we ensured that each session lasts for 10 minutes. During this time, we executed several port scans from both attackers, where each attacker is connected to a different base station. We also captured the benign traffic flows during the whole time. Table \ref{tab:timings} shows the timings of each session of port scans.

\begin{table*}[htb]
\caption{Schedule of the Port Scans}
    \label{tab:timings}
    \centering
    \begin{tabular}{|m{3.6cm}|m{1.4cm}|m{4cm}|m{4cm}|} 
    \hline
    Type of Port Scan & Date & Period of attack (Attacker 1) & Period of attack (Attacker 2) \\ \hline \hline
       SYN Scan & 26/08/2022 & 12.20pm-12.30pm & 12.20pm-12.30pm\\ \hline
       TCP Connect Scan & 26/08/2022 & 12.45pm-12.55pm & 12.45pm-12.55pm\\ \hline
       UDP Scan & 26/08/2022 & 2.45pm-3.15pm & 2.45pm-3.15pm\\ \hline
    \end{tabular}

\end{table*}

A data capturing session for each DoS attack lasted for 30 minutes. However, during a 30 minutes session, we executed DoS attack from each attacker only for 10 minutes. We scheduled the 10 minutes attacking period such that there is a 5 minutes period where both attackers perform DoS attacks simultaneously. We generated the benign traffic at all times throughout the 30 minutes period and captured the same. Table \ref{tab:dosattacks} shows the schedule of the DoS attacks.

\begin{table*}[htb]
\caption{Schedule of the DoS/DDoS Attacks}
    \label{tab:dosattacks}
    \centering
    \begin{tabular}{|m{3.6cm}|m{1.40cm}|m{2.9cm}|m{3.7cm}|m{3.7cm}|} 
    \hline
    Type of DoS/DDoS Attack & Date & Period of data collection & Period of attack (Base Station 1) & Period of attack (Base Station 2) \\ \hline \hline
       ICMP flood  & 26/08/2022 & 2.45pm-3.15pm & 2.50pm-3.00pm & 2.55pm-3.05pm \\ \hline
       UDP flood & 26/08/2022 & 3.50pm-4.20pm & 3.55pm-4.05pm & 4.00pm-4.10pm \\ \hline
       SYN flood & 29/08/2022 & 10.30am-11.00am & 10.35am-10.45am & 10.40am-10.50am \\ \hline
       HTTP flood & 29/08/2022 & 1.15pm-1.45pm & 1.20pm-1.30pm & 1.25pm-1.35pm\\ \hline
       Slow rate DoS - Slowloris & 29/08/2022 & 02.10pm-2.40pm & 2.15pm-2.25pm & 2.20pm-2.30pm \\ \hline
       Slow rate DoS - Torshammer & 29/08/2022 & 2.45pm-3.15am & 2.50pm-3.00pm & 2.55pm-3.05pm \\ \hline
    \end{tabular}
    
\end{table*}

\subsection{Data Processing}

\begin{figure}[ht]
    \centering
    \includegraphics[width=0.4\textwidth]{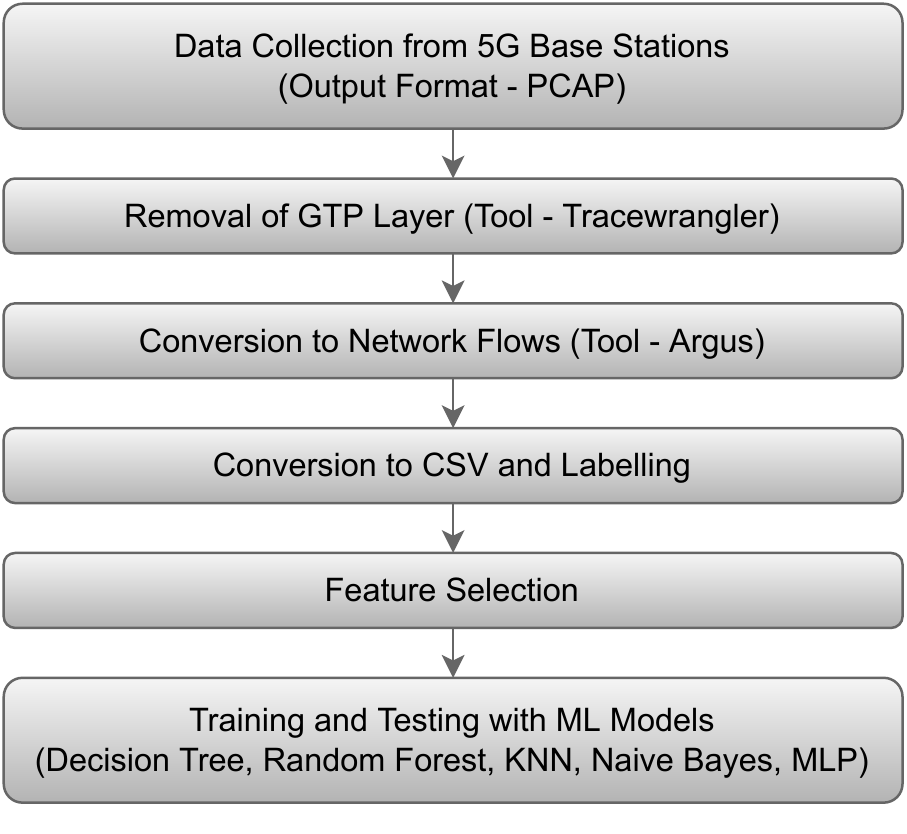}
    \caption{Data processing steps after collection.}
    \label{fig:dataprocessingsteps}
\end{figure}

We collected the data from each session for the two base stations into two pcap files. The duration of the overall captured traffic was 8 hours. The number of pcap files available was 18. Post processing of raw data was mandatory before feeding into the ML models. It is also important to obtain meaningful information and to increase the quality of the overall dataset. We performed a series of post-processing steps on the captured data as depicted in Fig.~\ref{fig:dataprocessingsteps}.

\subsubsection{GTP layer removal}

The capture of data from the radio interface of the base station is a novel approach which we do not see in other datasets. Due to this, the captured packets contained a GTP-U layer. GTP-U is a protocol used in IP based communications to transmit packets between the radio access network and the core network in Long-Term Evolution (LTE) or 5G New Radio (NR) networks. For the accurate identification of important features within the data, it was important to remove the GTP layer. We performed this process using the \textit{Tracewrangler} tool on a Windows PC. \textit{Tracewrangler} has the ability to remove the GTP layers in large batches of packets at once.

\subsubsection{Generating Network flows}

As the next step of the data processing, we converted the data from the packet-based format to a flow-based format. Intrusion detection systems consider network traffic in both the packet level as well as flow level. However, with the massive number of packets generated under different attack scenarios, the scanning process of each individual packet is time consuming and a bottleneck for the performance of the entire network~\cite{sourdis2006packet}. Moreover, modern attacks may span over multiple packets, therefore network flow based analysis provides better insights than the packet level analysis.

Hence, flow based formats are popular in ML based intrusion detection systems. A flow is a collection of packets with common characteristics such as the same source and destination IP addresses, same source, and destination ports as well as the same protocol~\cite{sperotto2010overview}. The flow based format provides a high level description of the common features between transmissions~\cite{li2013survey}. This conversion resulted in a shrinking of the total data volume, which is an additional advantage in terms of processing in the detection stage.

We converted the traffic from packet based format to flow based format through \textit{Argus} tool. \textit{Argus} is an open source tool that supports network flow data generation for analysis. When converting to network flows, we generated 112 features to distinguish each flow using \textit{Argus}. This included main features such as source and destination IP address, ports, protocol, packet size in bytes, number of packets.

\subsubsection{Aggregation of data and labelling}

We converted the flow data in argus format with 112 features into Comma Separated Values (CSV) format. The data collected at each attack session was available as separate CSV files. The data labeling process is an important step to ensure the quality of a dataset used for supervised learning. In this case, we labeled the data to differentiate between benign and attack traffic. We also assigned a different label for each attack.

After the labeling process, we aggregated the individual session data per base station. It makes sure that separate datasets are also available from multiple points of the network, if needed. Finally, we combined the two datasets from two base stations together to obtain a dataset with 1,215,890 flows.

\subsubsection{Data encoding}

We observed some categorical data in the 112 features extracted from the network flows. Since there is no meaning for categorical data in the context of ML, we converted them into numerical characters using a popular encoding technique; one hot encoding \cite{potdar2017comparative}. One hot encoding converts n observances in a particular variable with d distinct categories into d number of variables with n observations each. The n observations of these variables are represented with either 0 or 1 depending on the presence or absence of the particular category. We used one hot encoding in our own python script for this conversion. 

\subsection{Feature Selection}

Feature selection is an important step before feeding data into ML models. Modern massive heterogeneous networks generate a massive amount of data. The higher volume of data requires higher processing and storage requirements. Therefore, the selection and extraction of the most important features is a common practice in AI/ML model training. This also reduces the training times significantly. Further, the processing of high-dimensional data is challenging for most ML models. Feature selection removes irrelevant and redundant data effectively to save computational time and obtain higher learning efficiency~\cite{guyon2003introduction},~\cite{cai2018feature}.

Feature selection picks a subset called the most important features from an original set of features using techniques such as filter, wrapper, and embedded method. We used the filter method to conserve the generality, as data will be used for training different ML models. The filter method uses a statistical approach which is performed before applying to a model. Therefore, it is independent of the characteristics or parameters of a model \cite{stanczyk2015feature}. 

Before performing statistical-based filter methods for feature selection, we removed some obvious irrelevant features. Initially, we removed the features having null values and constant values throughout all the flows as they have no effect on the model accuracy. Then we removed certain features to maintain the generality of the intrusion detection system. These features include source and destination IP addresses and ports. We removed them because attack can be uniquely identified with their presence. For further reduction of features, we used statistical-based filter methods such as Pearson correlation and Chi-square test.

\subsubsection{Pearson Correlation}

From the remaining set of features, if more than one feature has the same relationship with the output label, then they are redundant. The dataset needs only one feature from those redundant features. We eliminated redundant features and further reduced the number of features without affecting the quality of data. We used the Pearson correlation coefficient for this. Pearson correlation coefficient is a statistical formulation used to determine the linear correlation between two variables based on~(\ref{eq1}). The correlation coefficient varies between -1 and +1 where +1 represents a high positive correlation and -1 represents a high negative correlation. A correlation value of 0 usually means that the two variables are uncorrelated.

\begin{equation}
\label{eq1}
\begin{split}
\rho_{x,y} = \frac{COV(X,Y)}{\sigma_x\sigma_y} = \frac{E(X-\mu_x)(Y-\mu_y)}{\sigma_x\sigma_y}
\end{split}
\end{equation}

\begin{figure}[hbt]
    \centering
    \includegraphics[width=8cm, height=9cm]{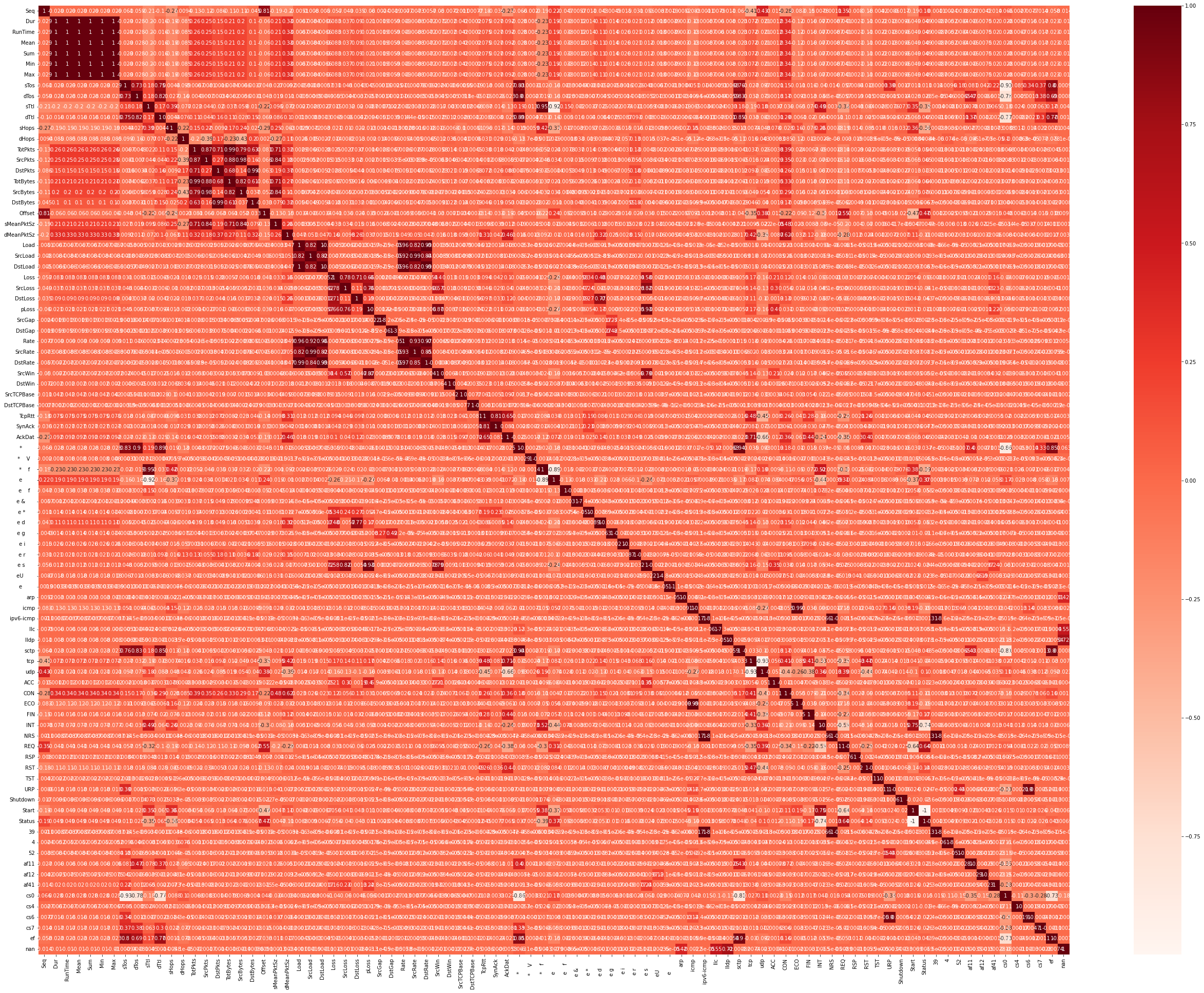}
    \caption{Heatmap with Pearson correlation between all features.}
    \label{fig:heatmap}
\end{figure}

We determined the pairwise correlation coefficient for every single pair of features. Fig.~\ref{fig:heatmap} shows the heatmap obtained using the Pearson correlation coefficient among all features. We considered the absolute values of the correlation coefficient since a high negative correlation coefficient also signifies a strong relationship. We chose a threshold of 0.90 to determine a high correlation among a pair of features and we removed one feature from such a pair. In this case, the eliminated feature out of a feature pair was the one with the least correlation to the target variable.

After the removal of all redundant features, we used the ANOVA F-value to determine the top features and we performed the same procedure for both binary and multiclass classifications.

\subsubsection{ANOVA (Analysis of Variance) F-scores}

Due to the presence of numerical data as input variables and the objective of categorizing the final output, we considered the ANOVA F-score as the ideal method to rank the top features~\cite{brownlee2019choose}. The ANOVA F-score is a feature selection method that ranks features based on ratio between variances~\cite{dissanayake2021comparative}. We used (\ref{eq2}) and (\ref{eq3}) for calculating the variance between the groups and within the groups respectively. Then we used the ratio of these two values to calculate the F-scores according to (\ref{eq4}). 

\begin{equation} \label{eq2}
\begin{split}
Variance\ between\ groups = \frac{ \sum_{i=1}^{n} n_i(\Bar{Y_i} - \Bar{Y})^2 }{(K-1)}
\end{split}
\end{equation}

\begin{equation} \label{eq3}
\begin{split}
Variance\ within\ groups = \frac{ \sum_{i=1}^{K} \sum_{j=1}^{n_i} (Y_{ij} - \Bar{Y})^2 }{(N-K)}
\end{split}
\end{equation}

\begin{equation} \label{eq4}
\begin{split}
ANOVA\ F-score = \frac{Variance\ between\ groups}{Variance\ within\ groups}
\end{split}
\end{equation}

where $\Bar{Y_i}$ is the sample mean of group $i$, $n_i$ is the number of observations in group $i$, $\Bar{Y}$ is the mean of the overall data, $K$ is the number of groups, $N$ is the total sample size, and $Y_{ij}$ is the observation $j$ in group $i$.

We implemented the ANOVA test using Scikit learn library and determined ANOVA F-scores between each feature and the target variable to eliminate irrelevant features. We obtained the top 25 features for both binary classification and multiclass classification. Table \ref{ANOVAbinary} and Table \ref{ANOVAmulti} respectively present the ANOVA F-scores for binary and multiclass classification with their scores.

\begin{table}[ht!]
    \caption{Ranking of Top 25 Features based on ANOVA F-scores for Binary Classification} 
    \label{ANOVAbinary} 
    \centering
    \begin{tabular}{|m{2cm}|c|} 
    \hline
    Feature  & Score \\ \hline \hline
    Seq	& 329,589.08 \\
	Offset	& 223,791.80 \\
	sTtl	& 189,934.06 \\
	e	& 159,060.91 \\
	tcp	& 142,813.28 \\
	AckDat	& 80,707.00 \\
	RST	& 35,379.19 \\
	INT	& 33,941.29 \\
	TcpRtt	& 32,500.88 \\
	icmp	& 29,713.08 \\
	sMeanPktSz	& 26,963.61 \\
	FIN	& 24,472.87 \\
	sHops	& 24,055.76 \\
	Mean	& 23,238.27 \\
 	dTtl	& 14,719.08 \\
	SrcBytes	& 12,836.57 \\
	TotBytes	& 11,701.19 \\
	dMeanPktSz	& 10,184.31 \\
	Status	& 9,574.53 \\
	cs0	& 7,384.06 \\
	SrcWin	& 6,928.11 \\
	sTos	& 6,341.94 \\
	e d	& 2,927.15 \\
	CON	& 2,668.01 \\
	REQ	& 750.70 \\
 \hline 
\end{tabular}
\end{table}

\begin{table}[ht!]
    \caption{Ranking of Top 25 Features based on ANOVA F-scores for Multiclass Classification} 
    \label{ANOVAmulti} 
    \centering
    \begin{tabular}{|m{2cm}|c|} 
    \hline
    Feature  & Score \\ \hline \hline
    tcp	& 1,460,107.76 \\
	AckDat & 283,300.90 \\
	sHops &	159,802.03 \\
	Seq &	52,291.33 \\
	RST &	50,894.45 \\
	TcpRtt &	47,017.48 \\
	REQ	& 44,521.06 \\
    dMeanPktSz    &	41,572.58 \\
	Offset &	38,844.09 \\
	CON	& 37,448.56 \\
	FIN	& 36,644.63 \\
	sTtl &	24,115.22 \\
	e	& 20,410.25 \\
	INT	& 17,248.98 \\
	Mean &	15,394.60 \\
	Status &	13,879.31 \\
	icmp &	8,376.06 \\
	SrcTCPBase &	6,592.89 \\
	e d	& 4,434.32 \\
	sMeanPktSz	& 4,183.54 \\
	DstLoss	& 2,625.58 \\
	Loss	& 1,954.14 \\
	dTtl	& 1,839.87 \\
	SrcBytes	& 1,606.26 \\
	TotBytes	& 1,587.40 \\
 \hline 
\end{tabular}
\end{table}

\subsection{Data normalization}

The reduction of training time is a valuable metric to determine the efficiency of any intrusion detection system. Data normalization also helps to reduce training times. Data normalization before feeding data into a ML model for training involves the adjustment of values to a predetermined common range so that larger values will not dominate the smaller values during model training~\cite{singh2020investigating}. We used the Z-score normalization technique in this work as stated in (\ref{eq5}). Z-score normalization uses the mean and the standard deviation of the dataset to re scale the data to have zero mean and unit variance~\cite{singh2020investigating}.

\begin{equation} \label{eq5}
\begin{split}
\hat{x}_{i,n} = \frac{x_{i,n} - \mu_i}{\sigma_i}
\end{split}
\end{equation}

\subsection{File description}

5G-NIDD contains data in both packet based format as well as in flow based formats. The data collected at each attack session from the two base stations are available separately in pcapng format. The same files with the removal of GTP layer are available in packet based pcapng format, argus and csv file formats. Argus and csv formats are flow based. We concatenated the GTP layer removed flow based csv files to produce files that contain all attacks from both base stations. The dataset contains csv files of all data collected from each base station as well as the combined dataset of base stations. Finally, the encoded version of the dataset is also available. This file does not contain unnecessary fields such as null fields. The dataset is available at http://ieee-dataport.org/10203. A summary of each of the files available in the dataset is shown in Table~\ref{filedescription}.

\begin{table*}[ht!]
\caption{Description of All Files in the Dataset}
 \label{filedescription} 
    \centering
\begin{tabular}{|m{3.6cm}|m{0.8cm}|m{12cm}|}
\hline
\textbf{File Name} & \textbf{Format} & \textbf{Description}\\ \hline \hline
Attackname\_BSX.pcapng & pcapng & Data collected from base station X on each attack session in packet-based format\\ \hline
Attackname\_BSX\_nogtp.pcapng & pcapng & Data collected from base station X on each attack session in packet-based format with GTP layer removed\\ \hline
AttacknameX.argus & argus & Further processed Attackname\_BSX\_nogtp.pcapng file to generate network flows \\ \hline
AttacknameX.csv & csv & Network flows in CSV format\\ \hline
BTS\_X.csv & csv & Concatenated data for all attack scenarios for base station X as network flows\\ \hline
Combined.csv & csv & Concatenated data for all attack scenarios for both base stations as network flows\\ \hline
Encoded.csv & csv & Network flows from both base stations containing all attack scenarios with categorical data encoded \\ \hline
\end{tabular}
\end{table*}

\section{Analysis and Discussion}
\label{sec:analysis}

This section provides details on the analysis performed on the dataset using different ML models including deep neural networks. It also describes the performance measuring metrics individually along with the ML models used. The section concludes with a comparison of performance for each ML model for both binary and multiclass classification.

\subsection{Performance measuring metrics}

We used the most common performance evaluation matrices such as confusion matrix, accuracy, precision, and F1-score to evaluate the performance and compare the results.

\subsubsection{Confusion Matrix}

A confusion matrix is a primary metric that derives all other metrics in classification problems. In a binary classification problem, the classified data are either True Positive (TP), False Positive (FP), True Negative (TN), or False Negative (FN). Table~\ref{tab:confusionmatrixbinary} outlines the confusion matrix for binary classification and Table~\ref{tab:confusionmatrixmulti} shows the confusion matrix for a multiclass classification. In both cases, the rows represent the predicted class whereas the columns represent the actual class. 

\begin{table}[ht]
\caption{Confusion Matrix for Binary Classification}
    \label{tab:confusionmatrixbinary}
    \centering
    \begin{tabular}{|m{2.8cm}|m{2.25cm}|m{2.27cm}|} 
    \hline
     & Actual Positive Class & Actual Negative Class \\ \hline \hline
     Predicted Positive Class & True Positive (TP) & False Negative (FN) \\ \hline
     Predicted Negative Class & False Positive (FP) & True Negative (TN) \\ \hline
    \end{tabular}
\end{table}

\begin{table}[ht]
\caption{Confusion Matrix for Multiclass Classification}
    \label{tab:confusionmatrixmulti}
    \centering
    \begin{tabular}{|m{1.5cm}|c|c|c|c|}
    \hline
    & \multicolumn{4}{c|}{Actual} \\ \hline \hline
    & Classes & Class 1 & Class 2 & Class 3 \\ \cline{2-5}
    \multirow{3}{*}{Predicted} & Class 1 & TP & $E_{21}$ & $E_{31}$ \\ 
     & Class 2 & $E_{12}$ & TP & $E_{32}$\\ 
     & Class 3 & $E_{13}$ & $E_{23}$ & TP\\ \hline
    \end{tabular}
\end{table}

\subsubsection{Accuracy}

The accuracy is the ratio of correctly classified samples to all the samples. It is the primary metric used in determining the performance of ML models. In a multiclass problem, the accuracy can be calculated per class, or for the entire model.

\begin{equation} \label{eq6}
\begin{split}
Accuracy = \frac{TP + TN}{TP + FP + TN + FN}
\end{split}
\end{equation}

\subsubsection{Precision}

Precision is the ratio of true positives to the total number of predicted positive values. In multiclass classifications, to calculate the precision of a given class, the true positives of that class and predicted positives of all the classes are taken into account.

\begin{equation} \label{eq7}
\begin{split}
Precision = \frac{TP}{TP + FP}
\end{split}
\end{equation}

\subsubsection{Recall}

Recall is the ratio of positive class which is correctly classified to the total number of actual positives. With more classes, this is calculated per class. In that case, the Recall of a given class is the ratio between the number of correct predictions for that class and the total number of actuals in that class.

\begin{equation} \label{eq8}
\begin{split}
Recall = \frac{TP}{TP + FN}
\end{split}
\end{equation}

\subsubsection{F1-score}

The f1-score is a weighted average or harmonic mean of both the precision and recall values. The precision and recall are inversely proportional to each other as the increase of one parameter leads to a decrease of the other. In such instances, the F1-score is a good measure to evaluate the performance of an ML model effectively.

\begin{equation} \label{eq9}
\begin{split}
F1-Score = \frac{2 * (Precision * Recall)}{Precision + Recall} 
\end{split}
\end{equation}

\subsection{ML Algorithms}

We considered different ML models to obtain the performance metrics to ensure the quality of the dataset. We also considered the training time and prediction times of each model in addition to the metrics previously mentioned. We tuned the hyperparameters using Grid search to select the most optimized parameters for each model before the execution of each model. A description of each model we used is given below.

\subsubsection{Decision Tree}

The decision tree algorithm is a common ML model in classification and regression problems. It classifies data samples based on specific rules. Decision trees have three components namely decision nodes, branches, and leaves. Each decision node recognizes a test attribute whereas a branch indicates the decision based on the value of the test attribute. Finally, the leaves represent the specific class that the attribute belongs to. We implemented the decision tree algorithm through Scikit learn library in Python.

\subsubsection{Random Forest}

The Random Forest algorithm is a collection of decision trees producing a result through a voting process. Each decision tree performs a class prediction and the class with the most votes is the final prediction class. The higher number of decision trees reduces the errors thereby increasing the accuracy levels. Specifically, the over-fitting problem in decision trees is solved by Random Forest, which also has a high tolerance for noise and outliers~\cite{tan2019wireless}.

\subsubsection{K-Nearest Neighbor}

The K-Nearest Neighbor (KNN) classification is another well-established classification model in supervised learning tasks. In KNN, an unlabeled data point is assigned a category identifier based on the category of the nearest neighbors to the particular data point. Several parameters such as the number of neighbors to be considered can be specified in the model inputs. KNN produces desirable results, especially in IDS, and a suitable benchmark for other classifiers based on accuracy levels~\cite{lin2015cann}.

\subsubsection{Naive Bayes}

The Naive Bayes classifier is a more general classification algorithm based on the Bayesian theory. According to Bayesian theory, the probabilities of future events can be calculated through their prior frequency of occurrence. The Naive Bayes classifier works on the assumption that the values of attributes are conditionally independent given a target value~\cite{islam2007investigating}. Although Naive Bayes is a simple classifier, it has the capability to perform very well with high dimensional input data~\cite{islam2007investigating}.

\begin{table*}[hbt]
\caption{Results of Different Evaluation Metrics obtained for Binary Classification with top 10 Features}
    \label{tab:binary10}
    \centering
    \begin{tabular}{|m{2.1cm}|m{1cm}|c|c|c|c|c|c|} 
    \hline
     Model & Type & Precision & Recall & F1-Score & Accuracy & Training Time (s) & Prediction Time (s) \\ \hline \hline
     \multirow{2}{*}{Decision Tree} & Benign & 0.99922 & 0.99809 & 0.99867 & \multirow{2}{*}{0.998955772} & \multirow{2}{*}{4.322835279} & \multirow{2}{*}{0.028925467} \\ 
      & Malicious & 0.99877 & 0.99951 & 0.99913 & & &
      \\ \hline
      \multirow{2}{*}{Random Forest} & Benign & 0.99921 & 0.99943 & 0.99934 & \multirow{2}{*}{0.999467331} & \multirow{2}{*}{8.763574076} & \multirow{2}{*}{0.221343637} \\ 
      & Malicious & 0.99964 & 0.99949 & 0.99954 & & &
      \\ \hline
        \multirow{2}{1pt}{KNN} & Benign & 0.99906 & 0.99756 & 0.99831 & \multirow{2}{*}{0.998675045} & \multirow{2}{*}{2.20724678} & \multirow{2}{*}{338.5157223} \\ 
      & Malicious & 0.99842 & 0.99940 & 0.99890 & & &
      \\ \hline
         \multirow{2}{1pt}{Naive Bayes} & Benign & 0.95937 & 0.94721 & 0.99831 & \multirow{2}{*}{0.963472299} & \multirow{2}{*}{1.241901278} & \multirow{2}{*}{0.064836335} \\ 
      & Malicious & 0.96606 & 0.97401 & 0.97004 & & &
      \\ \hline
         \multirow{2}{1pt}{MLP} & Benign & 0.99821 & 0.99800 & 0.99812 & \multirow{2}{*}{0.998520425} & \multirow{2}{*}{150.2028009} & \multirow{2}{*}{0.260591626} \\ 
      & Malicious & 0.99874 & 0.99883 & 0.99879 & & &
      \\ \hline
    \end{tabular}
\end{table*}

\subsubsection{Multi Layer Perceptron}

Multi Layer Perceptron (MLP) is a fully connected feed forward Artificial Neural Network (ANN) that is widely deployed as a sophisticated and versatile classification technique in IDS~\cite{catania2012automatic}. The design of an optimal MLP leads to better performance in speed and accuracy. The performance is mainly based on the number of hidden layers and the number of neurons in each layer~\cite{esmaily2015intrusion}. For our work, we used an MLP with 3 hidden layers having 10, 20, and 10 neurons in each layer to obtain optimal performance.

\subsection{Binary Classification}

We performed a binary classification on the dataset to distinguish the malicious flows from benign flows. We used a train test split of 70\% and 30\% out of the total flows. To determine the optimal number of features from the top 25 features, we conducted the experiments for 5, 10, 15, 20, and 25 of the topmost features obtained from the feature selection performed in~\ref{featureselect}. We achieved optimal performance in accuracy and computational time with the top 10 features. We repeated each experiment 10 times and used their average as the final value presented in this paper. Table~\ref{tab:binary10} presents the evaluation metrics obtained for each of the models. Tables~\ref{tab:confusionmatrixdt} to~\ref{tab:confusionmatrixmlp} show the confusion matrices obtained for each of the classification models. We executed the algorithms using Jupyter Notebook on a supercomputer known as CSC Puhti with hardware specifications; 4 CPU cores, 64 GB memory, and 20 GB local disk.

\begin{table}[H]
\caption{Confusion Matrix for Decision Tree at 10 Features}
    \label{tab:confusionmatrixdt}
    \centering
    \begin{tabular}{|c|c|c|} 
    \hline
     & True - Benign & True - Malicious \\ \hline \hline
     Predicted - Benign & 143,114 & 272 \\ \hline
     Predicted - Malicious & 109 & 221,278 \\ \hline
    \end{tabular}
\end{table}

\begin{table}[H]
\caption{Confusion Matrix for Random Forest at 10 Features}
    \label{tab:confusionmatrixrf}
    \centering
    \begin{tabular}{|c|c|c|} 
    \hline
     & True - Benign & True - Malicious \\ \hline \hline
     Predicted - Benign & 143,329 & 80 \\ \hline
     Predicted - Malicious & 114 & 221,245 \\ \hline
    \end{tabular}
\end{table}

\begin{table}[H]
\caption{Confusion Matrix for KNN at 10 Features}
    \label{tab:confusionmatrixknn}
    \centering
    \begin{tabular}{|c|c|c|} 
    \hline
     & True - Benign & True - Malicious \\ \hline \hline
     Predicted - Benign & 143,059 & 350 \\ \hline
     Predicted - Malicious & 134 & 221,219 \\ \hline
    \end{tabular}
\end{table}

\begin{table}[H]
\caption{Confusion Matrix for Naive Bayes at 10 Features}
    \label{tab:confusionmatrixnb}
    \centering
    \begin{tabular}{|c|c|c|}  
    \hline
     & True - Benign & True - Malicious \\ \hline \hline
     Predicted - Benign & 135,861  &  7,574 \\ \hline
     Predicted - Malicious & 5,751  & 215,572 \\ \hline
    \end{tabular}
\end{table}

\begin{table}[H]
\caption{Confusion Matrix for MLP at 10 Features}
    \label{tab:confusionmatrixmlp}
    \centering
    \begin{tabular}{|c|c|c|}  
    \hline
     & True - Benign & True - Malicious \\ \hline \hline
     Predicted - Benign & 143,124 & 285 \\ \hline
     Predicted - Malicious & 255 & 221,090 \\ \hline
    \end{tabular}
\end{table}

\subsection{Multi Class Classification}

We also performed a multiclass classification to validate the performance in detecting the attack type. We conducted this classification for 9 different attacks called HTTP flood, ICMP flood, SYN flood, SYN scan, Slowrate DoS, TCP connect scan, UDP flood, and UDP scan along with benign traffic flows. We used the same training test split of 70:30 in this classification too. Among experiments conducted with the top most 5, 10, 15, 20, and 25 features, the experiments with the top 10 features provided greater accuracy levels with acceptable training and prediction times. Table~\ref{tab:multi10} shows the average results of 10 executions of the experiment for each model. Tables~\ref{confusionmatrixdtmulti_dt} to~\ref{confusionmatrixdtmulti_mlp} present the confusion matrices for each model.

\begin{table*}[htbp]
\caption{Results of Different Evaluation Metrics obtained for Multiclass Classification with Top 10 Features}
    \label{tab:multi10}
    \centering
    \begin{tabular}{|m{1.5cm}|m{2.5cm}|c|c|c|c|c|c|} 
    \hline
     Model & Class & Precision & Recall & F1-Score & Accuracy & Training Time (s) & Prediction Time (s) \\ \hline \hline
      \multirow{9}{1pt}{Decision Tree} & Benign & 0.99309 & 0.99302 & 0.99306 & \multirow{9}{*}{0.991678249} & \multirow{9}{*}{3.880316758} & \multirow{9}{*}{0.038118291} \\ 
      {} & HTTP Flood & 0.98951 & 0.98712 & 0.98828 & & & \\
      {} & ICMP Flood & 0.97894 & 0.99820 & 0.98775 & & & \\
      {} & SYN Flood & 0.99649 & 0.98289 & 0.98956 & & & \\
      {} & SYN Scan & 0.98283 & 0.99700 & 0.98954 & & & \\
      {} & Slowrate DoS & 0.97575 & 0.97967 & 0.97760 & & & \\
      {} & TCP Connect Scan & 0.98609 & 0.95721 & 0.97122 & & & \\
      {} & UDP Flood & 0.99798 & 0.99598 & 0.99698 & & & \\
      {} & UDP Scan & 0.89662 & 0.96370 & 0.92714 & & & \\
      \hline
      \multirow{9}{1pt}{Random Forest} & Benign & 0.99503 & 0.99761 & 0.99633 & \multirow{9}{*}{0.994812031} & \multirow{9}{*}{11.7032125} & \multirow{9}{*}{0.37769897} \\ 
      {} & HTTP Flood & 0.98819 & 0.99281 & 0.99048 & & & \\
      {} & ICMP Flood & 1.00000 & 0.99943 & 0.99971 & & & \\
      {} & SYN Flood & 0.99702 & 0.99183 & 0.99439 & & & \\
      {} & SYN Scan & 0.99724 & 0.99695 & 0.99711 & & & \\
      {} & Slowrate DoS & 0.98643 & 0.97586 & 0.98108 & & & \\
      {} & TCP Connect Scan & 0.99417 & 0.97523 & 0.98451 & & & \\
      {} & UDP Flood & 0.99915 & 0.99724 & 0.9982 & & & \\
      {} & UDP Scan & 0.95747 & 0.96848 & 0.96273 & & & \\
      \hline
      \multirow{9}{1pt}{KNN} & Benign & 0.99666 & 0.99726 & 0.99698 & \multirow{9}{*}{0.994835607} & \multirow{9}{*}{2.148859835} & \multirow{9}{*}{282.0652506} \\ 
      {} & HTTP Flood & 0.98666 & 0.98946 & 0.98805 & & & \\
      {} & ICMP Flood & 0.99878 & 0.99943 & 0.99910 & & & \\
      {} & SYN Flood & 0.99436 & 0.99945 & 0.99690 & & & \\
      {} & SYN Scan & 0.99853 & 0.99637 & 0.99744 & & & \\
      {} & Slowrate DoS & 0.97945 & 0.97656 & 0.97799 & & & \\
      {} & TCP Connect Scan & 0.99547 & 0.98896 & 0.99220 & & & \\
      {} & UDP Flood & 0.99931 & 0.99774 & 0.99853 & & & \\
      {} & UDP Scan & 0.95534 & 0.97696 & 0.96588 & & & \\
      \hline
      \multirow{9}{1pt}{Naive Bayes} & Benign & 0.40038 & 0.98916 & 0.57077 & \multirow{9}{*}{0.409909614} & \multirow{9}{*}{1.292910719} & \multirow{9}{*}{0.227512336} \\ 
      {} & HTTP Flood & 0.49171 & 0.00988 & 0.01895 & & & \\
      {} & ICMP Flood & 0.00000 & 0.00000 & 0.00000 & & & \\
      {} & SYN Flood & 0.20000 & 0.00006 & 0.00014 & & & \\
      {} & SYN Scan & 0.10000 & 0.09939 & 0.09969 & & & \\
      {} & Slowrate DoS & 0.37849 & 0.00105 & 0.00208 & & & \\
      {} & TCP Connect Scan & 0.77147 & 0.99439 & 0.86869 & & & \\
      {} & UDP Flood & 0.00000 & 0.00000 & 0.00000 & & & \\
      {} & UDP Scan & 0.00000 & 0.00000 & 0.00000 & & & \\
      \hline
      \multirow{9}{1pt}{MLP} & Benign & 0.99650 & 0.99605 & 0.99630 & \multirow{9}{*}{0.994992694} & \multirow{9}{*}{316.8805176} & \multirow{9}{*}{0.230028152} \\ 
      {} & HTTP Flood & 0.98935 & 0.99345 & 0.99140 & & & \\
      {} & ICMP Flood & 0.99855 & 0.99855 & 0.99855 & & & \\
      {} & SYN Flood & 0.98990 & 0.99950 & 0.99465 & & & \\
      {} & SYN Scan & 0.99880 & 0.99545 & 0.99710 & & & \\
      {} & Slowrate DoS & 0.98705 & 0.97955 & 0.9832 & & & \\
      {} & TCP Connect Scan & 0.98775 & 0.99715 & 0.99240 & & & \\
      {} & UDP Flood & 0.99720 & 0.99850 & 0.99785 & & & \\
      {} & UDP Scan & 0.97995 & 0.94005 & 0.95955 & & & \\
      \hline
    \end{tabular}
\end{table*}

\begin{table*}[hbt]
\caption{Confusion Matrix for Decision Tree Multiclass Classification with 10 Features}
\centering
\label{confusionmatrixdtmulti_dt}
\begin{tabular}{|m{3.5cm}|p{1cm}|p{1cm}|p{1cm}|p{1cm}|p{1cm}|p{1cm}|p{1.2cm}|p{1cm}|p{1cm}|} \hline
              & True - Benign & True - HTTP flood & True - ICMP flood & True - SYN flood & True - SYN scan & True - Slowrate DoS & True - TCP connect scan & True - UDP flood & True - UDP scan \\ \hline \hline
Predicted - Benign & 142,409 & 4 & 2 & 5       & 96       & 2       & 53 & 248 & 565       \\ \hline
Predicted - HTTP flood & 7        & 41,780    & 0        & 0        & 0     & 538        & 0 & 0 & 0     \\ \hline
Predicted - ICMP flood & 1        & 0        & 347    & 0        & 0        & 0        & 0 & 0 & 0       \\ \hline
Predicted - SYN flood & 37        & 0      & 0        & 2,874    & 3        & 0        & 10 & 0 & 0        \\ \hline
Predicted - SYN scan & 4        & 0        & 0        & 1        & 5,961    & 0        & 5 & 0 & 8       \\ \hline
Predicted - Slowrate DoS & 7        & 436        & 0        & 0        & 0       & 21,428    & 2 & 0 & 0        \\ \hline
Predicted - TCP Connect Scan & 243        & 1        & 0       & 4        & 4        & 1        & 5,753 & 0 & 4   \\ \hline
Predicted - UDP flood & 544        & 0        & 7        & 0        & 0        & 0        & 0 & 136,632 & 0   \\ \hline
Predicted - UDP scan & 149       & 0        & 0        & 0        & 13        & 0        & 11 & 0 & 4,586   \\
\hline
\end{tabular}
\end{table*}

\begin{table*}[hbt]
\caption{Confusion Matrix for Random Forest Multiclass Classification with 10 Features}
\centering
\label{confusionmatrixdtmulti_RF}
\begin{tabular}{|m{3.5cm}|p{1cm}|p{1cm}|p{1cm}|p{1cm}|p{1cm}|p{1cm}|p{1.2cm}|p{1cm}|p{1cm}|} \hline
              & True - Benign & True - HTTP flood & True - ICMP flood & True - SYN flood & True - SYN scan & True - Slowrate DoS & True - TCP connect scan & True - UDP flood & True - UDP scan \\ \hline \hline
Predicted - Benign & 143,067     & 2       & 0        & 6        & 0       & 3       & 20 & 114 & 197       \\ \hline
Predicted - HTTP flood & 14        & 41,981     & 0        & 0        & 0     & 290        & 0 & 0 & 0     \\ \hline
Predicted - ICMP flood & 0        & 0        & 311    & 291        & 0        & 0        & 0  & 0 & 0       \\ \hline
Predicted - SYN flood & 16        & 0     & 0        & 2,318    & 1,187        & 0        & 9 & 0 & 2        \\ \hline
Predicted - SYN scan & 4        & 80        & 0        & 2        & 4,773    & 4,306        & 3 & 0 & 5       \\ \hline
Predicted - Slowrate DoS & 43      & 419        & 0        & 0        & 1       & 17,039    & 1,199 & 0 & 1        \\ \hline
Predicted - TCP Connect Scan & 149        & 2        & 0        & 0        & 4        & 1        & 4,663 & 27,390 & 4   \\ \hline
Predicted - UDP flood & 376        & 0        & 0        & 0        & 1        & 0        & 1 & 109,415 & 931   \\ \hline
Predicted - UDP scan & 139        & 0        & 0        & 0        & 12        & 0        & 3 & 0 & 4,597   \\
\hline
\end{tabular}
\end{table*}

\begin{table*}[hbt]
\caption{Confusion Matrix for KNN multiclass Classification with 10 Features}
\centering
\label{confusionmatrixdtmulti_KNN}
\begin{tabular}{|m{3.5cm}|p{1cm}|p{1cm}|p{1cm}|p{1cm}|p{1cm}|p{1cm}|p{1.2cm}|p{1cm}|p{1cm}|} \hline
              & True - Benign & True - HTTP flood & True - ICMP flood & True - SYN flood & True - SYN scan & True - Slowrate DoS & True - TCP connect scan & True - UDP flood & True - UDP scan \\ \hline \hline
Predicted - Benign & 143,016     & 50       & 0        & 11        & 1       & 7       & 24 & 94 & 206       \\ \hline
Predicted - HTTP flood & 8        & 41,840     & 0        & 0        & 0     & 437        & 0 & 0 & 0     \\ \hline
Predicted - ICMP flood & 0        & 0        & 347    & 0        & 0        & 0        & 0  & 0 & 0      \\ \hline
Predicted - SYN flood & 1        & 0      & 0        & 2,922    & 0        & 0        & 0   & 0 & 0     \\ \hline
Predicted - SYN scan & 6        & 0        & 0        & 5        & 5,957    & 0        & 2 & 0 & 7       \\ \hline
Predicted - Slowrate DoS & 4        & 510        & 0        & 0        & 0       & 21,539    & 0   & 0 & 0     \\ \hline
Predicted - TCP Connect Scan & 46        & 7        & 0        & 1       & 3        & 5        & 5,944 & 0 & 6   \\ \hline
Predicted - UDP flood & 310        & 0        & 0        & 0        & 0        & 0    & 0   & 136,873 & 0   \\ \hline
Predicted - UDP scan & 103        & 0        & 0        & 0        & 5        & 0        & 2 & 0 & 4,649   \\
\hline
\end{tabular}
\end{table*}

\begin{table*}[hbt]
\caption{Confusion Matrix for Naive Bayes Multiclass Classification with 10 Features}
\centering
\label{confusionmatrixdtmulti_NB}
\begin{tabular}{|m{3.5cm}|p{1cm}|p{1cm}|p{1cm}|p{1cm}|p{1cm}|p{1cm}|p{1.2cm}|p{1cm}|p{1cm}|} \hline
              & True - Benign & True - HTTP flood & True - ICMP flood & True - SYN flood & True - SYN scan & True - Slowrate DoS & True - TCP connect scan & True - UDP flood & True - UDP scan \\ \hline \hline
Predicted - Benign & 142,502     & 9       & 0        & 0        & 0   & 0    & 898       & 0 & 0       \\ \hline
Predicted - HTTP flood & 41,866        & 418     & 0        & 0        & 0     & 1        & 0 & 0 & 0     \\ \hline
Predicted - ICMP flood & 347        & 0        & 0    & 0        & 0        & 0        & 0 & 0 & 0        \\ \hline
Predicted - SYN flood & 2,510       & 0      & 0        & 0    & 0        & 0        & 413 & 0 & 0        \\ \hline
Predicted - SYN scan & 4,911        & 0        & 0        & 0        & 602    & 0        & 465 & 0 & 0       \\ \hline
Predicted - Slowrate DoS & 21,820        & 29        & 0        & 0        & 0       & 18    & 0  & 0 & 0      \\ \hline
Predicted - TCP Connect Scan & 34        & 0        & 0        & 0        & 0        & 0        & 5,977 & 0 & 0   \\ \hline
Predicted - UDP flood & 137,183        & 0        & 0        & 0        & 0        & 0        & 0 & 0 & 0   \\ \hline
Predicted - UDP scan & 4,758        & 0        & 0        & 0        & 0        & 0        & 0 & 0 & 0   \\
\hline
\end{tabular}
\end{table*}

\begin{table*}[hbt]
\caption{Confusion Matrix for MLP Multiclass Classification with 10 Features}
\centering
\label{confusionmatrixdtmulti_mlp}
\begin{tabular}{|m{3.5cm}|p{1cm}|p{1cm}|p{1cm}|p{1cm}|p{1cm}|p{1cm}|p{1.2cm}|p{1cm}|p{1cm}|} \hline
              & True - Benign & True - HTTP flood & True - ICMP flood & True - SYN flood & True - SYN scan & True - Slowrate DoS & True - TCP connect scan & True - UDP flood & True - UDP scan \\ \hline \hline
Predicted - Benign & 142,549     & 6       & 1        & 20       & 4      & 4       & 68 & 387 & 76      \\ \hline
Predicted - HTTP flood & 3       & 42,185     & 0        & 0        & 0     & 276        & 0 & 0 & 0     \\ \hline
Predicted - ICMP flood & 1        & 0        & 354    & 0        & 0        & 0        & 0     & 0 & 0   \\ \hline
Predicted - SYN flood & 1        & 0      & 0        & 2,890    & 0        & 1        & 0  & 0 & 0       \\ \hline
Predicted - SYN scan & 5        & 0        & 0        & 8       & 5,898    & 1        & 4 & 0 & 11       \\ \hline
Predicted - Slowrate DoS & 0        & 443        & 0        & 0        & 0       & 21,288    & 1 & 0 & 0        \\ \hline
Predicted - TCP Connect Scan & 1        & 6        & 0        & 2        & 1        & 2        & 5,943 & 1 & 6   \\ \hline
Predicted - UDP flood & 210        & 0        & 0        & 0        & 0        & 0        & 0 & 137,350 & 0   \\ \hline
Predicted - UDP scan & 283        & 0        & 0        & 0        & 3        & 0        & 1 & 0 & 4,486   \\
\hline
\end{tabular}
\end{table*}

\subsection{Analysis of Results}

The overall results we obtained for binary classification were very good for all the ML models except for the Naive Bayes model. The evaluation metrics such as Accuracy, Precision, Recall, and F1-Score were almost identical for Decision Tree, Random Forest, KNN, and MLP with slight variations. The Naive Bayes classification showed considerably lower performance in all the evaluation metrics. The confusion matrices obtained for each of the models show that Random Forest performs better in identifying malicious traffic with the least amount of false negatives. Moreover, the Decision Tree has been able to produce the lowest amount of false positives.

Different models showed different training and prediction times. The Naive Bayes model took the least amount of time to train, followed by the KNN and Decision Tree models. The Multi Layer Perceptron neural network consumed a relatively higher time for the training process than the other models. The prediction time was quite low for all the models except for KNN which required a considerably larger amount of time for prediction than training.

The accuracy levels obtained for multiclass classification were also very good for each attack. Except for the Naive Bayes model which performed rather poorly, all other models performed exceptionally well in detecting the attack types almost at similar accuracy levels. We can safely say that this model is not a suitable model for these types of problems. A closer look at the confusion matrix obtained for each of the models shows that different models performed well in detecting certain types of attacks among others. The overall accuracy MLP is marginally ahead of others. 

Generally among all models, the confusion matrices indicate that the classifier incorrectly classifies a considerable portion of UDP flood and UDP scan network flows as benign traffic compared with other flows. Furthermore, another common occurrence has been the mis-classification between HTTP flood and Slowrate DoS attack flows.

\section{Discussions About Potential Future Work and Further Usages of Testbed and Dataset}
\label{sec:futurework}
 
This paper describes our 5G testbed and highlights its capability and application for mobile/cellular security research. More specifically, this paper focuses on the testbed enabling data-driven study to secure the availability of mobile networks against DoS threats. This section discusses other potential future directions enabled by the testbed and the dataset.
 
\noindent \textbf{Distributed Detection:} 5G-NIDD uses two attack sources and two base stations. We can make it distributed in both the attack and defense (distributed defense against DDoS) by placing the attackers and the target in different parts of the network. This would enhance the possibility of testing federated learning models. We can further expand the defense implementations, i.e., defense implementation on a greater number of network nodes, to study distributed/joint defense and its effectiveness gain compared to a single point of defense for future work. The defense effectiveness can improve in the detection accuracy (enabled by greater information sensed and collected from multiple sources) and in mitigation (containing the DoS threat impact by implementing active defenses in greater nodes). Furthermore, the defense network can utilize distributed ML or federated learning.
 
\noindent \textbf{Active Defenses:} This paper focuses on detection and information processing, which is critical as it builds the intelligence for active defense control. Potential future studies can involve the active defenses when the DoS threats are detected and identified, including filtering, networking bandwidth/rate control, re-routing, and priority channel construction.

\noindent \textbf{More Threats:} This paper implements and simulates eight DoS threat types to demonstrate the usage of the 5G testbed. We can explore more threats and test them in our testbed, including the other threat methods for DoS (e.g., source-address spoofing, reflector threat, amplification threat) or more advanced zero-day threats against networking. With the increasing attention to zero-day detection~\cite{yang2021deep,kim2021zero}, in particular, our dataset can be utilized to evaluate new mechanisms designed for the detection of DoS variations.

\noindent \textbf{Dynamic Data Labeling:} We control the attacker generating the DoS traffic and therefore have the ground-truth knowledge about when the attack is executed and which attack types. Instead of utilizing such ground truth for the labeling of the datasets, we can explore the dynamic labeling based on the performance degradation on the base stations or on the other legitimate users whose availability is getting denied and disrupted due to the DoS threats.
 
\noindent \textbf{Adversarial Machine Learning:} We can simulate the attacker tampering with the training of the ML for detection, including dynamically changing its behavior and injecting noise-like or carefully crafted data. In addition, it is getting more crucial for intrusion detectors to have the capability of resistance against evasion attacks, often relying on adversarial ML 
tools to make subtle perturbations~\cite{zhang2022adversarial}.
Both the creation and detection of evasion attacks using adversarial ML techniques would be interesting and one of the potential areas for future explorations.

\section{Conclusions}
\label{sec:conclusions}

5G-NIDD contains a complete capture of the network traffic from a functional 5G network in the presence of different attack scenarios such as Port Scans and DoS attacks. In addition to the attack traffic, the dataset contains nonmalicious traffic from real users providing a more realistic dataset. The dataset is fully labeled specifying each flow to be malicious or benign. Moreover, each flow has a label that shows the attack type and the attacking tool used to aid multiclass classification. 

The dataset consists of 1,215,890 network flows, where each falls under a specific type of attack or benign traffic as depicted in Table \ref{tab:attacktypes}. The benign traffic comprises HTTP, HTTPS, SSH, and SFTP traffic. In addition to the dataset with a total number of flows, the dataset is also available with data collected from each individual base station separately. The availability of such separated datasets from different points in the network is potentially beneficial for federated learning research.

As shown in the performance metrics, the dataset produces remarkable results in binary and multiclass classification problems with many ML models. This emphasizes that the dataset can be used as a valid resource by the researchers to test novel algorithms related to multiclass classification problems.

\begin{table}[H]
\caption{Number of Flows with Each Attack Type}
    \label{tab:attacktypes}
    \centering
    \begin{tabular}{|c|c|}
    \hline
    Attack Type   &  Number of Flows\\ \hline \hline
       Benign & 477,737 \\ \hline
       HTTP Flood & 140,812 \\ \hline
       ICMP Flood & 1,155 \\ \hline
       SYN Flood & 9,721 \\ \hline
       SYN Scan & 20,043 \\ \hline
       Slowrate DoS & 73,124 \\ \hline
       TCP Connect Scan & 20,052 \\ \hline
       UDP Flood & 457,340 \\ \hline
       UDP Scan & 15,906 \\ \hline
    \end{tabular}
    
\end{table}

\section*{Acknowledgement}

This work is supported by the Academy of Finland in 6Genesis Flagship (grant no. 318927) project. This work is also supported by the Institute of Information \& communications Technology Planning \& Evaluation~(IITP) grant funded by the Korea government~(MSIT) (No. 2021-0-02107, Collaborative research on element Technologies for 6G Security-by-Design and standardization-based International cooperation). This work has been performed utilizing the 5GTN infrastructure and resources at the Centre for Wireless Communications, University of Oulu.

\bibliographystyle{IEEEtran}
\bibliography{references}

\vfill

\end{document}